\documentclass[12pt,preprint2]{aastex}
\usepackage{emulateapj5}
\usepackage{onecolfloat5}

\newcommand{\expnt}[2]{\ensuremath{#1 \times 10^{#2}}}   
\newcommand{\gsim}{\gtrsim}
\newcommand{\lsim}{\lesssim}

\newcommand{\rxj}{RX~J0720.4$-$3125}
\newcommand{\rxjw}{RX~J1856.5$-$3754}
\newcommand{\cxo}{\textit{CXO}}
\newcommand{\chandra}{\textit{Chandra}}
\newcommand{\rosat}{\textit{ROSAT}}

\newcommand{\sax}{\textit{BeppoSAX}}
\newcommand{\xmm}{\textit{XMM}}

\newcommand{\rbs}{RX~J1308.8+2127}
\newcommand{\axp}{1E~2259+58.6}
\newcommand{\psr}{PSR~J1814$-$1744}
\newcommand{\oldpsr}{PSR~J2144$-$3944}

\begin{document}

\shorttitle{X-ray Timing of \rxj}
\shortauthors{Kaplan et al.}
\slugcomment{Accepted for publication in ApJL}

\twocolumn[
\title{X-ray Timing of the Enigmatic Neutron Star \rxj}
\author{D.~L.~Kaplan, S.~R.~Kulkarni}
\affil{Department of Astronomy, 105-24 California Institute of
Technology, Pasadena, CA 91125, USA}
\email{dlk@astro.caltech.edu, srk@astro.caltech.edu}
\author{M.~H.~van~Kerkwijk}
\affil{Sterrenkundig Instituut, Universiteit Utrecht,
Postbus 80000, 3508 TA Utrecht, The Netherlands} 
\email{M.H.vanKerkwijk@phys.uu.nl}
\author{\and H.~L.~Marshall}
\affil{Center for Space Research, Massachusetts Institute of Technology,
         Cambridge, MA 02139, USA}
\email{hermanm@space.mit.edu}

\begin{abstract}
\rxj\ is the third brightest neutron star in the soft X-ray sky and
has been a source of mystery since its discovery, as its long 8-s
period separates it from the population of typical radio pulsars.
Three models were proposed for this source: a neutron star accreting
from the interstellar medium, an off-beam radio pulsar, or an old,
cooling magnetar.  Using data from \chandra, \rosat, and \sax\ we are
able to place an upper limit to the period derivative, $|\dot{P}| <
\expnt{3.6}{-13}\mbox{ s s}^{-1}$ (3-$\sigma$).  While our upper limit
on $\dot P$ allows for the accretion model, this model is increasingly
untenable for another similar but better studied neutron star,
RX~J1856.5$-$3754, and we therefore consider the accretion model
unlikely for \rxj.  We constrain the initial magnetic field of \rxj\
to be $\lsim 10^{14}$~G based on cooling models, suggesting that it is
not and never was a magnetar, but is instead middle-aged neutron star.
We propose that it is either a long-period high-magnetic field pulsar
with $\dot P\sim 10^{-13}\mbox{ s s}^{-1}$ similar to \psr, or a
neutron star born with an initial period of $\approx 8.3$~s and $\dot
P\sim 10^{-15}\mbox{ s s}^{-1}$.  The proximity of \rxj\ is strongly
suggestive of a large population of such objects; if so, radio pulsar
surveys must have missed many of these sources.

\end{abstract}
]

\section{Introduction}
\label{sec:introduction}
\rxj\ was discovered by \citet{hmb+97} as a soft ($kT \sim 80$~eV),
bright X-ray source in the \rosat\ All-Sky Survey.  Given its very low
hydrogen column density ($N_H\sim \expnt{1}{20}\mbox{ cm}^{-2}$),
nearly sinusoidal 8.39-s pulsations, relatively constant X-ray flux,
and very faint ($B=26.6$~mag), blue optical counterpart
\citep{kvk98,mh98}, it was classified as a nearby, isolated neutron
star.

As one of the closest ($d\sim 300$~pc; \citealt*{kvka02}) neutron
stars, \rxj\ occupies a central position in our study of these
objects. However, the long period is puzzling, and has led to three
models: an old, weakly magnetized neutron star accreting matter from
the interstellar medium \citep{w97,kp97}; a middle-aged pulsar with
$\sim 10^{12}$~G magnetic field whose radio beams are directed away
from the Earth \citep{kvk98}; or an old magnetar (neutron star with
magnetic field $> 10^{14}$~G; \citealt{dt92}) that is kept warm by the
decay of its strong magnetic field \citep{hh98b,hk98}.  These models
predict different period derivatives: $\dot{P} < 5\times10^{-15}\mbox{
s s}^{-1}$, $\dot{P} \sim 10^{-15}$--$10^{-13}\mbox{ s s}^{-1}$, and
$\dot{P} \gsim {\rm few} \times 10^{-13}\mbox{ s s}^{-1}$,
respectively.

Motivated thus, we undertook timing observations of \rxj\ using the
\textit{Chandra X-ray Observatory} (\cxo), supplemented with
analysis of archival data from \rosat\ and \sax.

After submission of this paper, we became aware of  the work of \cite{zhc+02}
reporting a timing analysis of \rxj.  A
``Notes added in manuscript'' section regarding that analysis can
be found at the end of the manuscript.

\begin{deluxetable}{l l r r r r c c l r}
\tablecaption{Summary of Observations\label{tab:sum}}
\tablecolumns{10} 
\tablewidth{0pc} 
\tabletypesize{\scriptsize}
\tablehead{\colhead{Date} & \colhead{MJD} & \colhead{Exp.} &
\colhead{Span} & \colhead{Counts} & \colhead{BG} & \colhead{Facility}
& \colhead{Instrument/} & \colhead{Period\tablenotemark{c}} &
\colhead{${\rm TOA}-\mbox{MJD }50000$\tablenotemark{d}}\\ 
&\colhead{(day)} & \colhead{(ksec)} & \colhead{(ksec)} & &
\colhead{Counts\tablenotemark{a}} & & \colhead{Mode\tablenotemark{b}}
& \colhead{(s)} & \colhead{(TDB day)} \\ } 
\startdata 
1993-Sep-27& 49257.2 & 3.2 & 12.0& 5800 & 22.8 & \rosat & PSPC & 8.3914(4) &
$-$742.745297(3)\\ 
1996-Nov-03& 50390.9 & 33.7 & 65.7 & 12662 & 79.0
& \rosat & HRI & 8.39113(6)& 391.300750(2)\\
 1997-Mar-16 & 50523.1 &
18.1 & 99.4 & 407 & 15.4 & \textit{BeppoSAX} & LECS & 8.39103(9) &523.705635(4)\\
 1998-Apr-20 & 50923.2 & 8.1 &460.3 & 3074 & 17.1 &\rosat & HRI &8.391114(14)& 925.688213(5)\\
 2000-Feb-01 & 51575.3 & 5.4
& 305.5 &929& 1.3& \cxo & HRC-S+LETG 0 & 8.39111(2)\tablenotemark{e} &1577.039569(2) \\
 & & & &671& 127.0 & \cxo & HRC-S+LETG $\pm 1$& \nodata& \nodata \\
 2000-Feb-02 & 51576.1 & 26.3 & \nodata &4584 & 5.2& \cxo & HRC-S+LETG 0 & \nodata & \nodata\\
 & & & &3027& 454.0 &\cxo & HRC-S+LETG $\pm 1$& \nodata& \nodata \\
 2000-Feb-04 & 51578.7 & 6.1 & \nodata &1119 & 1.2& \cxo & HRC-S+LETG 0& \nodata& \nodata\\
 & && &687 &119.5 & \cxo & HRC-S+LETG $\pm 1$ & \nodata& \nodata\\
2001-Dec-04 & 52247.7 & 15.0 & 168.6& 31746 &229.8& \cxo & ACIS-S3/CC
& 8.391119(12)\tablenotemark{e} & 2248.6768200(8)\\
 2001-Dec-05 &
52248.2 & 10.6 & \nodata& 22825 &155.8& \cxo & ACIS-S3/CC & \nodata &
\nodata\\
 2001-Dec-06 & 52249.6 & 4.1 & \nodata & 8786 &61.4& \cxo &
ACIS-S3/CC & \nodata& \nodata\\
\enddata 
\tablenotetext{a}{Background
counts scaled to the source extraction area. $^{\rm b}$HRC-S+LETG 0
indicates order 0; HRC-S+LETG $\pm 1$ indicates orders $\pm 1$.
$^{\rm c}$Values in parentheses are 1-$\sigma$ errors in the last
decimal digit. $^{\rm d}$TOA is defined as the
maximum of the folded lightcurve nearest the middle of the
observation, as determined from the best-fit sine wave.  The ACIS/CC
times were corrected for spacecraft motion following \citet{zpst00}.
$^{\rm e}$All pointings for each of the \chandra\ HRC-S and \chandra\
ACIS datasets were processed together.}
\end{deluxetable}

\section{Observations}
The primary data consist of two sets of observations obtained from
\chandra: one using the HRC in the spectroscopic mode (HRC-S) with the
Low Energy Transmission Grating (LETG), and one using ACIS in the
continuous clocking (CC) mode. The primary and archival datasets are
summarized in Table~\ref{tab:sum}.

We processed the HRC-S data using the standard
pipeline\footnote{\url{http://asc.harvard.edu/ciao/threads/spectra\_letghrcs/}}
and extracted $0^{\rm th}$ order events from a circle with radius
10~pixels ($1\farcs3$). For the $\pm 1^{\rm st}$ orders, we extracted
events from a region $0.0006\degr$ wide in the cross-dispersion
direction (the \texttt{tg\_d} coordinate) and from $0.08\degr$ to
$0.35\degr$ along the dispersion direction (the \texttt{tg\_r}
coordinate). We extracted events from the ACIS data within
$\pm1\arcsec$ of the source.  We then used the \texttt{axBary} program
to barycenter the events in both these datasets.

The best fit position for \rxj, found by averaging the $0^{\rm th}$
order data from the three HRC-S datasets, is (J2000) $\alpha=07^{\rm
h}20^{\rm m}24\fs96$, $\delta=-31\degr25\arcmin49\farcs6$, with rms
uncertainty of $\approx 0\farcs6$ in each coordinate due to \cxo\
aspect uncertainties.  This is consistent ($1\farcs4$ away) with the
optical position \citep{kvk98}.  The X-ray source appears unresolved
and its profile is consistent with that of a point source (half-power
radius of $\approx 0\farcs5$).

For the \rosat\ HRI data, we extracted the events within a circle of
radius 45~pixels ($22\farcs5$) centered on the source.  We used a
circle of radius 200~pixels ($100\arcsec$) for the PSPC data.  These
events were barycentered using the \texttt{ftools} programs
\texttt{abc} and \texttt{bct} and corrected to Barycentric Dynamical
Time (TDB) according to \citet[][p.\ 14]{allen}.

We extracted the \sax\ LECS events within a circle with radius of
25~pixels ($200\arcsec$) centered on the source and restricted to
those with pulse-invariant (PI) amplitudes that were less than 90
(energies $< 0.95$~keV), in order to maximize the
signal-to-noise. Finally, we barycentered the events with the
\texttt{SAXDAS} tool \texttt{baryconv}.

\section{Timing Analysis}
\label{sec:timing}
\begin{figure}[ht!]
\plotone{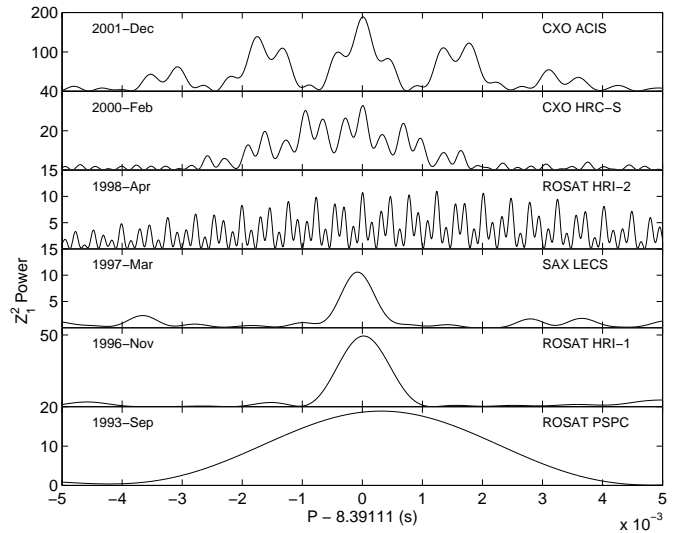}
\caption{$Z_{1}^{2}$ periodograms for the datasets listed in
Table~\ref{tab:sum}. For each dataset, the $Z_1^2$ power is
normalized so as to have unit mean (when no signal is present). Given
that the statistics of $Z_1^2$ are exponential it follows that the
variance is also unity.  Note the different vertical scales which
reflect the differing significance levels of the detections.  }
\label{fig:zn2}
\end{figure}

\begin{figure}[hb!]
\plotone{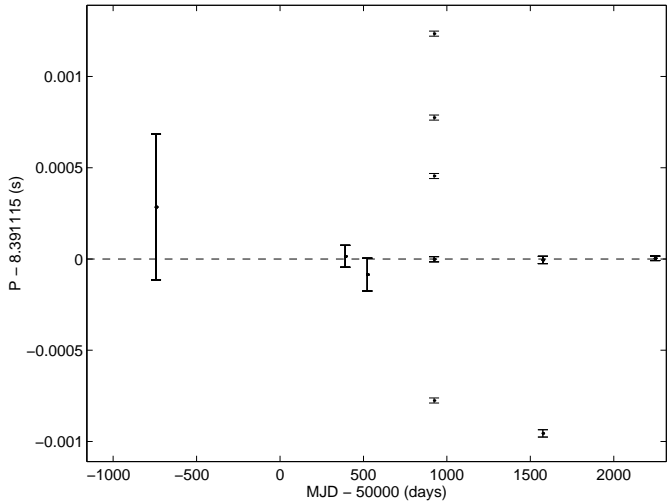}
\caption{Period measurements for \rxj, using the data from
Figure~\ref{fig:zn2}.  As explained in the main
text the \chandra\ HRC-S and \rosat\ HRI-2
measurements are ambiguous owing
to large gaps in the data. Probable periods are displayed. 
The best-fit constant period model is shown by the dashed line:
$P=8.391115(8)$~s.} 
\label{fig:sum}
\end{figure}

For each dataset, we computed $Z_{1}^{2}$ power spectra around the
known 8.39-s period.  Specifically, we explored the period range from
8.376~s to 8.405~s in steps of $7 \mbox{ }\mu{\rm s}$ (oversampling by
factors of 20--800 relative the nominal step-size of $P_{0}^{2}/\Delta
T$, where $P_{0}=8.39$~s is the approximate period and $\Delta T$ is
the span of the dataset from Table~\ref{tab:sum}).  As can be seen
from Figure~\ref{fig:zn2} all but the \chandra\ HRC-S and \rosat\
HRI-2 datasets yielded unambiguous period estimates. For the HRC-S and
HRI-2 sets the period estimates are ambiguous because the large gaps
in the observations result in strong side-lobes. In
Figure~\ref{fig:sum}, we display the best-fit periods for the
unambiguous determinations as well as possible periods for the HRC-S
and HRI-2 datasets.

As can be seen from Figure~\ref{fig:sum}, the ambiguity of the HRC-S
and HRI-2 datasets can be resolved provided we assume (reasonably)
that the period evolves smoothly with time. Our choice of period (for
HRC-S and HRI-2) and the best fit periods (for the other datasets) are
shown in Table~\ref{tab:sum}.  The errors on the periods were
determined using the analytical expression from \citet{ransom01}.
While that expression was derived for FFT power spectra, $Z_{1}^{2}$
power spectra have the same statistics (both are exponentially
distributed) so the same relations should apply (we have verified this
with numerical simulations).  We also show in Table~\ref{tab:sum}
times-of-arrival (TOAs) for each of the datasets.

The data in Table~\ref{tab:sum} are consistent with there being no
measurable $\dot{P}$: fitting for a linear spin-down gives
$P=8.391115(8)$~s at MJD~51633 and $\dot{P}=\expnt{(1 \pm
12)}{-14}\mbox{ s s}^{-1}$, with $\chi^{2}=1.54$ for 4
degrees-of-freedom (DOF).  If, instead, we fit only for a constant
period, we find $P=8.391115(8)$~s, with $\chi^{2}=1.64$ for 5 DOF.
Therefore we can constrain the secular period derivative to be $|
\dot{P} | < \expnt{3.6}{-13}\mbox{ s s}^{-1}$ (3-$\sigma$).

\begin{figure}[ht]
\plotone{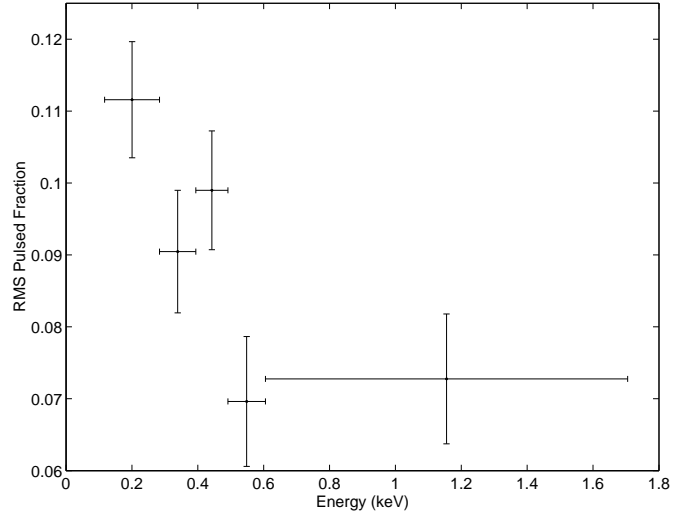}
\caption{RMS pulsed-fraction (see \S\ \ref{sec:timing} for
discussion) for different energy bins, from the
\chandra\ ACIS data.  The overall pulsed fraction is $8.1 \pm
0.4$\%. Each bin was chosen to have the same number of 
total counts.
}
\label{fig:pf}
\end{figure}

The folded lightcurve is largely sinusoidal, with an rms
pulsed-fraction (the rms of the lightcurve divided by the mean) of 8\%
for both \chandra\ datasets.  However, this pulsed-fraction is
energy-dependent: the fraction rises with decreasing energy (see
Figure~\ref{fig:pf}), in agreement with the \xmm\ analysis
\citep{pmm+01}.

\section{Discussion}
Our upper limit of $|\dot P| < \expnt{3.6}{-13}\mbox{ s s}^{-1}$ is
sufficiently high that we cannot meaningfully constrain the accretion
model, for which we expect $\dot P<\expnt{5}{-15}\mbox{ s s}^{-1}$
(the limit is the case where all of the required $\dot M$ of $\sim
10^{12}\mbox{ g s}^{-1}$ couples to the neutron star at the corotation
radius, giving maximum torque per unit mass).  We note that the
accretion model is no longer viable for another similar but better
studied isolated neutron star,
\rxjw\ \citep{vkk01b}.  Regardless, the accretion model is best
confronted by measuring the proper motion and distance, and looking
for evidence of sufficiently dense ambient gas (deep Keck H$\alpha$
imaging and \textit{HST} astrometric observations are in progress).

However, we can constrain the pulsar and magnetar models. We can draw
four inferences common to both models.  First, the spin-down
luminosity, $\dot E=I\dot\omega\omega < 2.4\times 10^{31}\,$ erg
s$^{-1}$; here, $\omega=2\pi/P$.  Second, in the framework of a simple
(vacuum magnetic dipole radiation) pulsar model, the physical age is
roughly approximated (provided the current spin period is much larger
than that at birth and that the magnetic field does not decay
significantly) by the so-called characteristic age: $\tau_c \equiv
P/(2\dot P) > \expnt{4}{5}$~yr.  Third, the strength of
the dipole field is $B=3.2\times 10^{19} (P\dot P)^{1/2} < 6\times
10^{13}$~G.  Fourth, we assume that the X-ray emission (well described
by a blackbody; \citealt{hmb+97,pmm+01}) is
cooling flux from the surface.  The bolometric
cooling luminosity is $L_{\rm cool} \approx
\expnt{2}{32}d_{300}^{2}\mbox{ ergs s}^{-1}$ \citep{hmb+97}, using the
distance estimate of $300 d_{300}$~pc derived by scaling from \rxjw\
\citep{kvka02}.

Knowledge of $L_{\rm cool}$ enables us to estimate the cooling age,
$t_{\rm cool}$, of \rxj. Magnetic fields, especially intense $B$
fields such as those proposed for magnetars, can profoundly influence
the cooling of neutron stars. To this end, we use the curves of $L$
vs.\ $t$ from \citet{hk98} and find $t_{\rm cool} \approx
\expnt{(5-10)}{5}$~yr,   assuming a 50\% uncertainty in the distance
and with only a slight dependence on $B$.  This age
is consistent with the characteristic age derived above.

In the magnetar model (we assume that the $B$-field decay is
dominated by the slower irrotational mode; see \citealt{hk98}) the
expected $B$ field  at about $10^6$~yr is, $\approx \expnt{2}{14}$~G, well above the
upper limit obtained from our $\dot P$ limit.  Models that are
consistent with both $L_{\rm cool}$ and our limit on $\dot P$ (and
thus an upper limit on the current value of $B$) are those with
initial $B \lsim 10^{14}$~G.  Based on this, we conclude that \rxj\ is
not a magnetar, motivating us to consider the pulsar model.

\begin{figure}[ht!]
\plotone{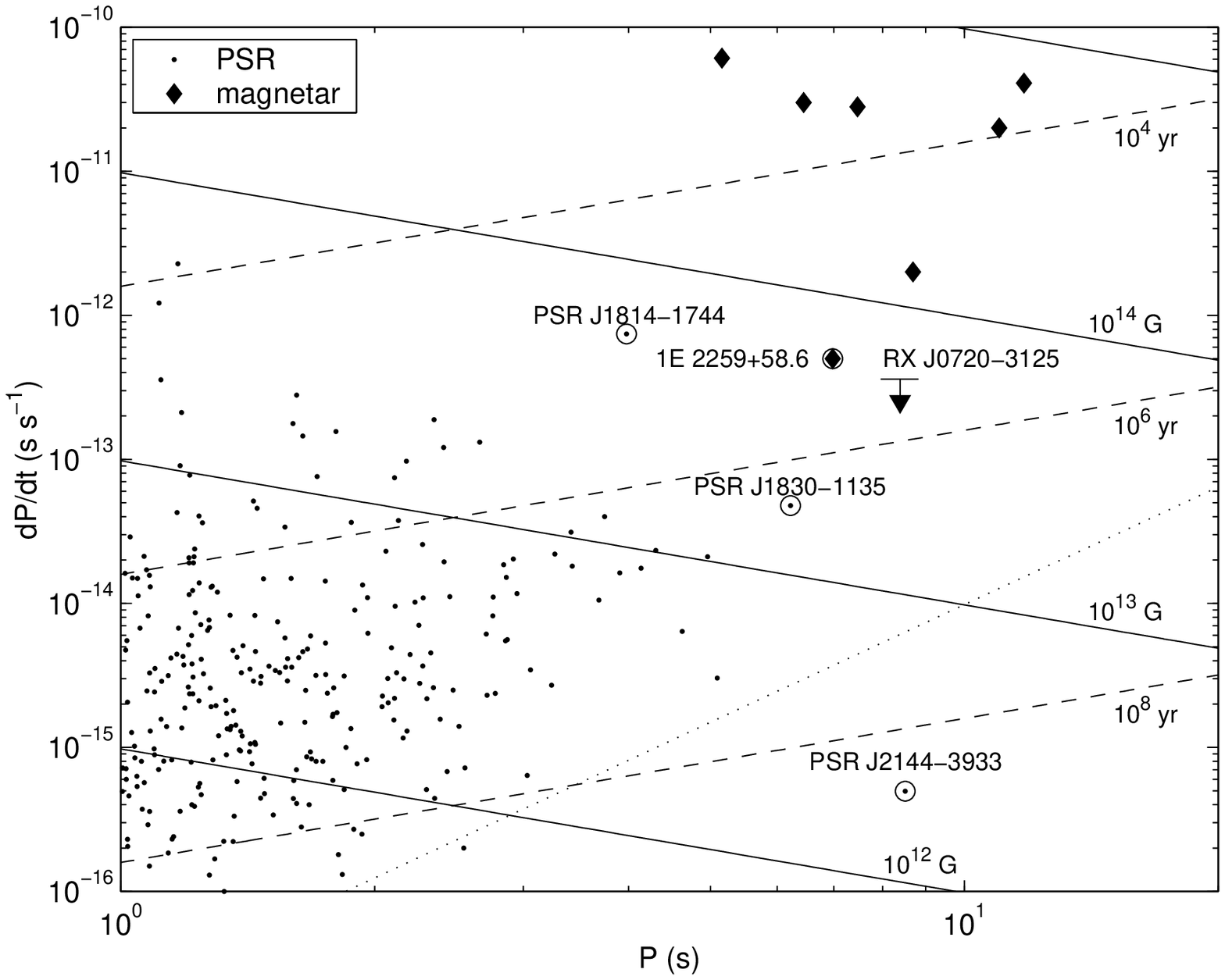}
\caption{$P$-$\dot{P}$ diagram, showing only $P\geq 1$~s and $\dot{P}
\geq 10^{-16}\mbox{ s s}^{-1}$.  Radio
pulsars  are plotted as points, magnetars as diamonds. 
\rxj\ is an upper limit.  The magnetar \axp\ is 
circled, as are the high-$B$ pulsars PSR~J1830$-$1135 and \psr, and 
the long-period pulsar \oldpsr\ \citep{ymj99}. A  version of the
so-called ``death line'' is marked 
by the dotted line.  
The sloping solid lines are lines of constant dipole magnetic field $B_{\rm
dipole} \equiv \expnt{3.2}{19}(P \dot{P})^{1/2}$~G, while the dashed
lines are those of constant characteristic age $\tau_{c} \equiv
P/(2\dot P)$.
\label{fig:ppdot}
}
\end{figure}

Earlier, \citet{kvk98} did not accept the radio pulsar model because
in 1998 there were no radio pulsars with such long periods.  However,
over the past four years we have come to appreciate the existence of
pulsars with $B > 10^{13}$~G (\citealt{ckl+00}; see also
Figure~\ref{fig:ppdot}).  In particular, the parameters of \rxj\ are
not too dissimilar to those of \psr, which has $P\approx 4$~s and
$\dot P\approx 7.4\times 10^{-13}\mbox{ s s}^{-1}$.  Thus, the past
objections against the radio pulsar model are no longer tenable, and
\rxj\ seems fully compatible with being an off-beam high-$B$ pulsar.  If that is
the case, then we expect $\dot{P} \sim 10^{-13}\mbox{ s s}^{-1}$, a
value that we should be able to measure in the near future.

A separate possibility is that \rxj\ is an off-beam pulsar with
age compatible with $t_{\rm cool}$, but with a
conventional ($\sim 10^{12}$~G) magnetic field and $\dot P \sim
10^{-15}\mbox{ s s}^{-1}$ (and therefore $\tau_{c}\sim
10^{8}$~yr). \rxj\ could then be similar to the 8.5-s,
$\expnt{2}{12}$-G pulsar \oldpsr\ (\citealt*{ymj99}; see
Figure~\ref{fig:ppdot}).  With a braking index of 3, the age of a
pulsar is $\tau = \tau_{c} \left( 1-(P_{0}/P)^{2}\right)$, where
$P_{0}$ is the initial spin-period.  If \rxj\ does have $\tau \sim
t_{\rm cool}$ and $B \sim 10^{12}$~G, we find and $P_{0}\approx
8.3$~s, very close to $P$.  Such as a pulsar would be an example of
the ``injection'' hypothesis \citep{vn81}, where pulsars are born with
initial spin periods $P_{0} \gg 10$~ms (as for the Crab).  Such long
initial periods are allowed and perhaps expected in some models of
neutron-star formation \citep[e.g.,][]{sp98}, where the precise
initial period depends very sensitively on the details of the
formation mechanism and may range over four orders of magnitude.
While there are a few pulsars whose characteristic ages are factors of
10--100 times the ages derived from supernova remnant associations
\citep{pzst02}, this would be the first case for a source with
$P_{0}>1$~s.

We make the following parenthetical observation: for most known
pulsars the X-ray pulsed-fraction (largely) increases with photon
energy \citep*{phh01}, whereas for \rxj\ we see the opposite
effect. However, for the pulsars (e.g.,\ PSR~B0656+14,
PSR~B1055$-$52), the X-ray luminosity has a strong, highly-pulsed,
non-thermal component with $L_{X,\mbox{\scriptsize non-th}} \sim
10^{-3}\dot{E}$ \citep{bt97}.  Furthermore, in such objects heating of
the polar caps by pulsar activity (probably dependent on $\dot E$) is
likely significant.  The interplay of these  components with
the viewing geometry can result in the large range of observed
phenomena \citep[e.g.,][]{phh01}.  For \rxj, though, with its small $\dot E$ there is little
reason to expect a strong non-thermal contribution or a hot polar cap
(although there must be some inhomogeneities to give the observed
pulsations). We conjecture that the increase in the pulse fraction
with decreasing photon energy is primarily due to the absence the
additional components.

\section{Conclusions}
\label{sec:conc}
In this {\it Letter}, based on X-ray timing data and cooling models,
we argue that the nearby soft X-ray source \rxj\ is not a middle-aged
magnetar but is likely a $10^{6}$-yr off-beam pulsar.  To accommodate
its age and long period we speculate that it either has $B\gsim
10^{13}$~G or was born with $P_{0}\approx 8.3$~s, a very surprising
result as both source types are, at present, considered to be rare.
We now consider the larger ramifications of our conclusions.
 
A volume-limited sample of neutron stars offers us an opportunity to
sample the diversity of such sources. In this respect, soft X-ray
surveys provide the best such samples since all neutron stars ---
normal radio pulsars, high-$B$ pulsars, magnetars, and the mysterious
Cas-A-like neutron stars --- will cool through soft X-ray emission
well into their middle ages.  Indeed, this expectation is borne out by
the local sample: pulsars such as PSR~B0656+14,
Geminga\footnote{Presumably a standard pulsar that is not beamed
toward us.}, \rxj, a youngish magnetar (see below), and finally the
very mysterious \rxjw, of which we know nothing other than it is a
cooling neutron star\footnote{It is further worth noting that the
sample of soft X-ray neutron stars has at least three long-period
objects \citep*{hpm99}.}.

The number of neutron stars belonging to a given class depends not
only on the sensitivity of the X-ray survey but also on the product of
the birth rate and the cooling age. Thus, for example, magnetars with
their longer-lasting cooling radiation may dominate the local
population despite a lower birthrate \citep{hk98}. This bias and the
long period led us to speculate that \rxj\ was an old magnetar, a
conclusion we have now refuted.  In contrast, the soft thermal X-ray
source \rbs, with $P=5.2\,$s and $\dot P\sim 10^{-11}\mbox{ s s}^{-1}$
\citep{hhss02} appears to have a magnetar-strength field.

The proximity of \rxj\ argues for a substantial Galactic population of
similar sources, but very few such radio pulsars are known.  The cause
of this paucity is that radio surveys select against long-period
pulsars, especially those with $B>10^{13}$~G, in several ways.  (1)
The beaming factor is known to decrease with increasing period,
reaching 3\% at $P\sim 10$~s \citep{tm98}.  (2) As can be seen from
Figure~\ref{fig:ppdot}, the lifetime of a radio pulsar decreases with
increasing $B$: a $B\sim 10^{12}$~G neutron star crosses the the radio
death line at $\sim 10^8$~yr whereas a $B\sim 10^{13}$~G pulsar dies
at $\sim 2\times 10^7$~yr. The loss of throughput of a pulsar survey
for a 5-s pulsar relative to a 1-s pulsar from these two effects alone
is nearly one order of magnitude.  (3) The true loss is even greater
since long-period signals are frequently classified as interference
(we note that population models do not constrain the
population of long-period pulsars [\citealt{hbwv97}], mainly due to
reasons 1 and 2).  Young high-$B$ pulsars in supernova remnants would
almost certainly create visible plerion nebulae due to their high
$\dot E$'s, while long-period injected pulsars of similar ages would
be invisible except for their cooling radiation (without assuming that
the radio beams are directed toward the Earth).  Thus, injected
pulsars detectable only via X-ray emission may be present in many
``hollow'' supernova remnants (i.e.\ those without visible plerions).
Radio pulsar searches better tuned to long periods and very deep radio
and X-ray searches for young pulsars in supernova remnants may uncover
the postulated class of long-period sources.

\acknowledgements We thank M.~Cropper and S.~Zane for discussions on
the TOAs and for alerting us to
the different time systems used by different programs.  We thank
D.~Frail and J.~Heyl for useful discussions, and we thank an anonymous
referee for constructive comments.  We have used the NASA-maintained
HEASARC web site for archival data retrieval and subsequent analysis.
D.L.K.\ holds a fellowship from the Fannie and John Hertz Foundation,
and his and S.R.K.'s research are supported by NSF and NASA. M.H.v.K.\
is supported by a fellowship from the Royal Netherlands Academy of
Arts and Sciences.

\textit{Notes Added In Manuscript}
After we submitted our paper we became aware of a paper by
\citet{zhc+02} reporting timing analysis of \rxj. Our period
determinations of the archival data (PSPC, HRI-1, LECS, HRI-2, HRC-S)
are in excellent agreement with those of Zane et al.  Both papers also
report new determinations, which are: \textit{Chandra} ACIS-S3 (our
paper) and XMMa (2000 May 13) and XMMb (2000 November 21; both from
Zane et al.).  We restricted our analysis to an incoherent combination
of the various datasets, i.e.\ we looked for secular evolution of the
period determined from each observation separately. We did not attempt
to phase connect the datasets.  Zane et al.\ do present a coherent
analysis, using the archival and \xmm\ data.  However, in our opinion
such an analysis is premature and not robust. First, it is premature,
because the \xmm\ derived periods of known pulsars have fractional
errors $|\Delta P/P|$ ranging from $1.9\times 10^{-7}$ to $1.2\times
10^{-5}$ (as reported by the \xmm\ calibration team; \citealt{kkb+02}).
This error alone may result in systematic uncertainties as high as
$\dot P$ of $6\times 10^{-12}$ over the 6-month duration of the \xmm\
datasets and $5\times 10^{-13}\,$s s$^{-1}$ over the entire span of
the observations.  Second, it is not robust, as phase connection without any
ambiguity requires that the datasets be separated by time intervals
less than ``coherence'' timescale, $\sim P^2/\sigma_{P}$ (where
$\sigma_{P}$ is the uncertainty in the measurement of $P$) and none of
the datasets (including the \xmm\ datasets) satisfy this condition.
We note that neither of the two primary solutions from Zane et al.\
fits the TOAs  in Table~\ref{tab:sum}.

\bibliographystyle{apj}


\end{document}